\documentclass[%
reprint,
amsmath,amssymb,
aps,
]{revtex4-1}

\usepackage{graphicx}
\usepackage{dcolumn}
\usepackage{bm}
\usepackage{braket}
\usepackage{amsmath}

\begin{document}


    \title{Study of the effect of newly calculated phase space factor on $\beta$-decay half-lives}
    \author{Jameel-Un Nabi$^{1}$, Mavra Ishfaq$^{1}$}
    \affiliation{$^{1}$GIK Institute of Engineering Sciences and Technology,
    Topi 23640, Khyber Pakhtunkhwa, Pakistan.}
    \email{jameel@giki.edu.pk}
    \author{$^{2, 3, 4}$Ovidiu Ni\c{t}escu,$^{3, 4}$Mihail Mirea, $^{3, 4}$Sabin Stoica}
    \affiliation{$^2$University of Bucharest, Faculty of Physics, P.O. Box MG11, 077125-Magurele, Romania}
    \affiliation{$^3$National Institute of Physics and Nuclear Engineering, P.O. Box MG6, 077125-Magurele, Romania}
    \affiliation{$^4$ International Centre for Advanced Training and Research in Physics, P.O. MG12, 077125-Magurele, Romania.}
    \email{sabin.stoica@cifra.infim.ro}
    \date{\today}

    \begin{abstract}

We present  results for $\beta$-decay half-lives based on a new
 recipe for calculation of phase space factors  recently introduced.
 Our study includes $fp$-shell and heavier nuclei of experimental and
 astrophysical interests. The investigation of the kinematics of
 some $\beta$-decay half-lives is presented, and new phase space factor values
 are compared with those obtained with previous theoretical approximations.
 Accurate calculation of nuclear matrix elements is a
 pre-requisite for reliable computation of $\beta$-decay
 half-lives and is not the subject of this paper. This paper
 explores if improvements in calculating the $\beta$-decay half-lives can be obtained when
 using a given set of nuclear matrix elements and employing the
 new values of the phase space factors.
 Although the largest uncertainty in half-lives computations
 come from the nuclear matrix elements,
 introduction of the new values of the phase space factors may
 improve the comparison with experiment.  The new  half-lives are
 systematically larger than previous calculations and may have
 interesting consequences for calculation of stellar rates.

        \begin{description}
            \item[PACS numbers]23.40.Bw; 23.40.-s;  26.30.Jk

        \end{description}
    \end{abstract}

    \pacs{Valid PACS appear here}

    \maketitle


    \section{\label{sec:level1}INTRODUCTION}
The precise knowledge of the $\beta$-decay rates represents an
important ingredient for understanding the nuclear structure as well
as the astrophysical processes like presupernova evolution of
massive stars, nucleosynthesis (s-, p-, r-, rp-) processes, etc.
\cite{Mol03,Ni14,Ren14}). That is why the calculation of the
$\beta$-decay half-lives in agreement with experimental results has
been a challenging problem for nuclear theorists
\cite{Zhi13,Niu13,Mar07,Hir93,Sta90}. Theoretically, the half-life
formulas for $\beta$-decay can be expressed as a product of nuclear
matrix elements (NMEs), involving the nuclear structure of the
decaying parent and of the daughter nuclei, and the phase space
factors (PSFs) that take into account the distortion of the electron
wave function by the nuclear Coulomb field. Hence, for a precise
calculation of the $\beta$-decay half-lives, an accurate computation
of both these quantities is needed. The largest uncertainties come
from the NME computation. In literature one can find different
calculations of the NMEs for $\beta$-decays, realized for different
types of transitions and final states, and with different
theoretical models (e.g. based on gross theory \cite{Tak75}, QRPA
approaches
\cite{Sta90,Hir93,Nab99,Wan16,Ni14,Nik05,Mar07,Lia08,Niu13,Tan17}
and shell model \cite{Mar99}). We would not be discussing
calculation of NMEs in this paper. Until recently the PSFs were
considered to be calculated with enough precision and, consequently,
not much attention was paid to a more rigorous calculation of them.
However, recently we recomputed the PSFs for positron decay and
electron capture (EC) processes for 28 nuclei of astrophysical
interest, using a numerical approach \cite{Sab16}. We
solved the Dirac equation (getting exact electron wave functions)
with a nuclear potential derived from a realistic proton density
distribution in the nucleus. We also included the screening effects.
The new recipe for
calculation can easily be extended to any arbitrarily heavy nuclei.

Accurate estimates of half-lives of neutron-rich nuclei have gained
much interest in the recent past. This is primarily because of their
key role in $r$-process nucleosynthesis. Similarly, precise value of
$\beta$-decay half-lives of proton-rich nuclei is a pre-requisite
for solving many astrophysical problems. In this paper, we study the
effect of introducing the new PSF values, obtained with our recently
introduced recipe \cite{Sab16} on the calculation of $\beta$-decay
half-lives. We also extend here our previous PSF calculations (of
positron decay and EC reactions) to include
$\beta$-decay reactions. In order to complete the calculation of
$\beta$-decay half-lives, we calculate the set of NMEs using the
proton-neutron quasi-particle random phase approximation model in
deformed basis  and a schematic separable potential both in
particle-particle and particle-hole channels. Other nuclear models
and a set of improved input parameters may result in a better
calculation of NMEs. However this improvement is not under the scope
of current paper. We calculate both Gamow-Teller and Fermi
transitions to ground and excited states, for medium and heavy
nuclei of interest. We present first an investigation of the
kinematics of $\beta$-decay half-lives and our PSF values are
compared with those obtained with previous theoretical
approximations. Later the newly computed half-lives are compared
with other previous theoretical predictions and experimental data.
We investigate if our new PSF values lead to any improvement in the
calculated $\beta$-decay half-life values. Our present study may be
extended to investigate the effect of the new PSF values on stellar
decay rates, which we take as a future assignment.

This paper is organized in the following format. Section II
describes the essential formalism for the calculation of  PSFs and
$\beta$-decay half-lives. We present our results in Section III
where we also make a comparison of the current calculations with
experimental data and previous calculation \cite{Gov71}. We conclude
finally in Section IV.

    \section{\label{sec:level2}FORMALISM}

    \subsection{Half Life Calculation}

$\beta$-decay half-lives can be calculated as a sum over all
transition probabilities to the daughter nucleus states through
excitation energies lying within the Q$_{\beta}$ value

\begin{equation}
T_{1/2} =( \sum_{0\leq E_{f}\leq Q_{\beta}} 1/t_{f})^{-1}
\label{hl},
\end{equation}
where the partial half-lives (PHL), $t_{f}$, can be calculated using

\begin{equation}
t_{f} = \frac{C}{(g_{A}/g_{v})^{2}F_A(Z, A, E)B_{GT}(E_{f})+
    F_V(Z,A,E)B_F(E_f)} \label{phl}.
\end{equation}

In Eq.~(\ref{phl}) value of C was taken as 6143 s \cite{Har09},
$g_{A}$, $g_{v}$ are axial-vector and vector coupling constants of
the weak interaction, respectively, having $g_{A}$/$g_{v}$= -1.2694
\cite{Nak10}, while $E_f$ is the final state energy. $E = Q_{\beta}
- E_f$ where $Q_{\beta}$ is the window accessible to either
$\beta^{+}$-, $\beta^{-}$- or EC decay.  $F_{A/V}$ are the PSFs.
$B_{GT}$ and $B_F$ are the reduced transition probabilities for
Gamow-Teller and Fermi transitions, respectively, and expressed as

\begin{equation}
B_{F}(E_f) = \frac{1}{2I_{i}+1} \mid<f \parallel M_{F}
\parallel i> \mid ^{2}, \label{ftp}
\end{equation}

\begin{equation}
B_{GT}(E_{f}) =\frac{1}{2I_{i}+1} \mid<f\parallel M_{GT} \parallel
i>\mid^{2} \label{rtp}
\end{equation}

In Eq.~(\ref{ftp}) and Eq.~(\ref{rtp}), $I_{i}$ denotes the spin of
the parent state, $M_{F}$ and $M_{GT}$ are the Fermi and
Gamow-Teller transition operators, respectively. Detailed
calculation of the NMEs within the proton-neutron quasi-particle
random phase approximation (pn-QRPA) formalism may be found in Refs.
\cite{Hir93,Sta90}.

In this paper the NMEs calculation was performed using the pn-QRPA
model. We used the Nilsson model \cite{Nil55} to calculate single
particle energies and wave functions which takes into account the
nuclear deformation. Pairing correlations were tackled using the BCS
approach. We considered proton-neutron residual interaction in two
channels, namely the particle-particle and the particle-hole
interactions. Separable forms were chosen for these interactions and
were characterized by interaction constants $\chi$ for
particle-particle and $\kappa$ for particle-hole interactions. Here,
we used the same range for $\chi$ and $\kappa$ as was discussed in
\cite{Hir93, Sta90}.  Deformation parameter values $\beta_{2}$ for all cases were
taken from Ref. \cite{Ram01}. For pairing gaps we used a global
approach $\Delta_{n}$ = $\Delta_{p}$ = 12/$\sqrt{A}$ [MeV]. A large
model space up to 7$\hbar\omega$ was incorporated in our model to
perform half-lives calculations for heavy nuclei considered in this
paper.

\subsection{Phase Space Factors Calculation}
\subsubsection{Phase space factors for $\beta^+/\beta^-$ transitions}
The formalism for the PSF calculation for $\beta^+/\beta^-$ allowed
transitions was discussed in detail in our previous paper \cite{Sab16}.
Here, we reproduce the main features of
the formalism for the sake of completion. The probability per unit
time that a nucleus with atomic mass A and charge Z decays for an
allowed $\beta$-branch is given by

\begin{eqnarray}\label{pro}
\lambda_{0} = G_{\beta}^{2}/ 2\pi^{3}\int_1^{W_0}
pW(W_0-W)^2S_{0}(Z, W)dW,
\end{eqnarray}
where $G_{\beta}$ is the weak interaction coupling constant, $p$ is
the momentum of $\beta$-particle, $W$ = $\sqrt {{p^{2} + 1}}$ is the
total energy of $\beta$-particle and $W_0$ is the maximum
$\beta$-particle energy. $W_0$ = $Q - 1$ ($Q + 1$) in $\beta^{+}$
($\beta^{-}$) decay. $Q$ is the mass difference between initial and
final states of neutral atoms. Eq.~(\ref{pro}) is written in natural
units ( $\hbar = m = c = 1$ ), so that the unit of momentum is $mc$,
the unit of energy is $mc^{2}$, and the unit of time is $\hbar$
/$mc^{2}$. The shape factors $S_{0}(Z, W)$ for allowed transitions
which appear in Eq.~(\ref{pro}) are defined as

\begin{eqnarray}\label{sf}
S_{0}(Z, W) = \lambda_1 (Z,W) |M_{0,1}|^2,
\end{eqnarray}
where $M_{0,1}$ are the NMEs related to the Fermi and Gamow-Teller reduced transition probabilities as

\begin{eqnarray}\label{nme}
|M_{0,1}|^{2} = \frac{1}{\sqrt {2I_{i}+1}} B_{F,GT},
\end{eqnarray}
and $\lambda_1 (Z,W)$ stands for Fermi functions. For the
calculation of the $\beta$-decay rates, one needs to calculate the
NMEs and the PSFs that can be defined as
\begin{equation}\label{ps1}
F_{\beta^+/\beta^-}=\int_1^{W_0} pW(W_0-W)^2\lambda_1(W)dW.
\end{equation}

The above formula determines the PSFs for both the Fermi and
Gamow-Teller allowed transitions, by substituting $F_V$ or $F_A$ in
Eq. (\ref{phl}), respectively. For the allowed $\beta$-decays the
Fermi functions can be expressed as
\begin{eqnarray}\label{bc}
\lambda_{1}(Z,W) = {g^2_{-1} + f^2_{1}\over 2p^2},
\end{eqnarray}
\noindent which is just the definition used in Ref. \cite{Gov71}
(Eq. (3)), for our particular case $k=1$.
 We note that in the above formula a coefficient $1/(2p^2)$ appears. Usually, this coefficient
is included in the proper normalization of the wave functions, as we
did. The functions $g_{-1}(Z,W)$ and $f_1(Z,W)$ are the large and
small radial components of the positron (or electron) radial wave
functions evaluated at the nuclear radius $R$. They are solutions of
the coupled set of differential equations \cite{Gov71}

\begin{eqnarray}
\label{dirac} \left(\frac{d}{dr} +
\frac{\kappa+1}{r}\right)g_{\kappa}(W,r) =
(W+V(r)+1)f_{\kappa}(W,r) \nonumber \\
\left(\frac{d}{dr} + \frac{\kappa-1}{r}\right)f_{\kappa}(W,r)=
-(W+V(r)-1)g_{\kappa}(W,r)
\end{eqnarray}
where $V(r)$ is the central potential for the positron (or the
electron) and $\kappa=(l-j)(2j+1)$ is the relativistic quantum
number.

\begin{figure}[h]
    \includegraphics [width=3.5in]{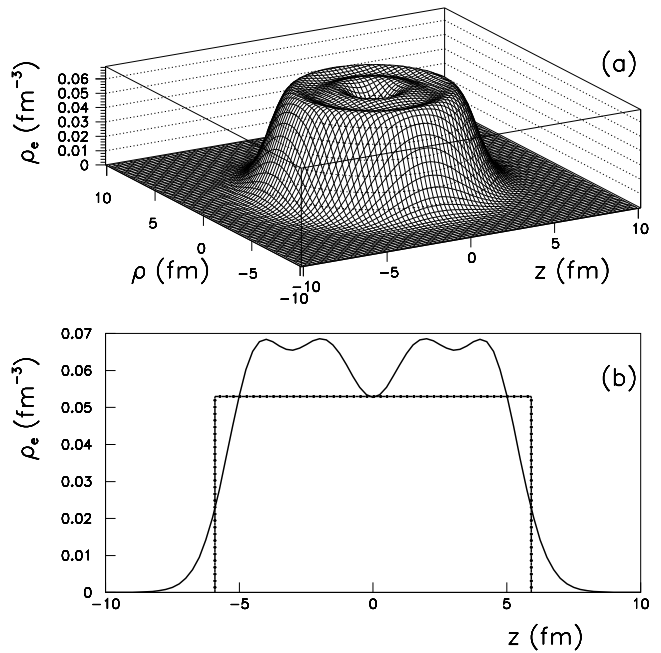}
    \centering \caption{\scriptsize (a) Realistic proton density for $ ^{120} $Xe
    represented in cylindrical coordinates. (b) Profile of the
    realistic proton density for $ ^{120} $Xe (thick line) compared
    with that given by a constant density approximation (dot-dashed line).  }
    \label{Xe_dens}
\end{figure}

Ideally, the central potential $V(r)$ from Eq.~(\ref{dirac}) should
include the effects of the extended nuclear charge distribution and
of the screening by orbital electrons. Unlike the recipe of Gove and
Martin \cite{Gov71}, where these screening effects were treated as corrections to the wave functions, in our recipe they are included directly in the
potential. This was done  by deriving the potential $V(r)$ from a
realistic proton density distribution in the nucleus. The charge
density can be written as

\begin{equation}\label{dens}
\rho_e(\vec{r})=\sum_{i}^{}(2j_i+1)v^2_i\left|\psi_i(\vec{r})\right|^2,
\end{equation}
where $\psi_i$ is the proton wave function of the spherical single
particle state $i$ and $v_i$ is its occupation amplitude. The wave
functions ${\psi_i}$, were found by solving Schr\"odinger equation
with a Wood-Saxon potential. The $(2j_i+1)$ term in Eq.~(\ref{dens})
reflects the spin degeneracy. As an example, we depict the realistic
proton density for $ ^{120} $Xe in cylindrical coordinates in
Fig.~\ref{Xe_dens}(a).  The profile of this proton density for the
daughter nucleus $^{120}$Xe (thick line) is compared with a constant
density (dot-dashed line) in Fig.~\ref{Xe_dens}(b).

\begin{figure}[h]
\includegraphics [width=3.5in]{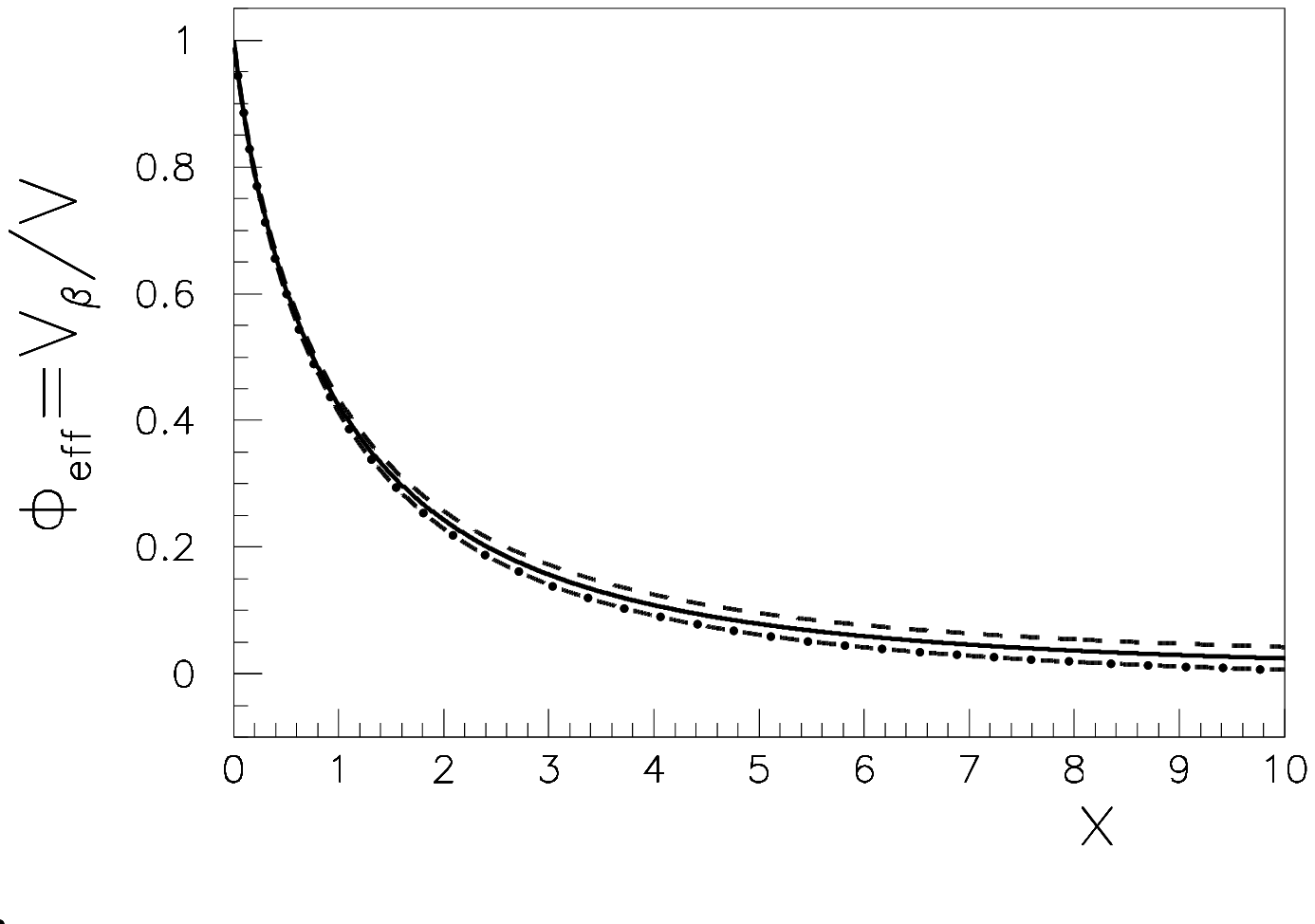}
    \centering \caption{ \scriptsize The effective screening function
$\phi_{eff}=V_{\beta^{\mp}}/V$ as function of the dimensionless
distance $x$ for $Z$=54. The dashed and the dot-dashed lines
correspond to $\beta^-$ and $\beta^+$ decays, respectively. The full
line gives the behavior of the screening function with the
boundaries $\phi(0)$=1 and $\phi(\infty)$=0, as given in Ref.
\cite{Esp02}.}
    \label{screening}
\end{figure}

We integrated the realistic charge distribution over the volume of
nucleus, in order to find the Coulomb potential

\begin{equation}\label{potential}
    V(Z,r)=\alpha\hbar c \int
    \frac{\rho_e(\vec{r'})}{|\vec{r}-\vec{r'}|}d\vec{r'}.
\end{equation}

Moreover, we included the screening effect by multiplying the
expression of $V(r)$ with a function $\phi(r)$, which is solution of
the Thomas-Fermi equation

\begin{equation}\label{Thomas-Fermi}
    \frac{d^2\phi}{dx^2}=\frac{\phi^{3/2}}{\sqrt{x}},
\end{equation}
with $x=r/b$, $b\approx0.8853a_0Z^{-1/3}$ and $a_0$ is the Bohr
radius. The solution $\phi(r)$ was calculated within the Majorana
method \cite{Esp02}. In the case of $\beta^- / \beta^+$, the
effective potential $V_{\beta^{\mp}}$ was modified by the screening
function $\phi(r)$ as

\begin{equation}\label{Vmodified}
    rV_{\beta^{\mp}}(Z,r)=(r V(Z,r)+ 1) \times \phi(r) - 1.
\end{equation}
So, asymptotically we returned the interaction between an ion of
charge $\pm 1$ of the residual nucleus, and the emitted
electron/proton. The Coulomb potential in atomic units is $Z/r$ and
is negative/positive for $\beta^-/\beta^+$ decays. As mentioned in
Ref. \cite{Esp02}, the Thomas-Fermi equation (\ref{Thomas-Fermi}) is
a universal equation which does not depend on $Z$ or other physical
constants. The boundaries of the screening function are $\phi(0)=1$
and $\phi(\infty)$=0. Then, the effective potential is
$V_{\beta^{\mp}}(r)=V(r)$ for $r$=0, and is suppressed according to
the variation of the universal screening function when $r$ increases
in order to reach the asymptotic behavior $V_{\beta^{\mp}}(r)=-1/r$
for $r\rightarrow\infty$. The effective screening function is
displayed in Fig. \ref{screening}. Such a procedure of including the
screening effect was also used previously in the computation of PSFs
for double-beta decays \cite{KI13}-\cite{MPS15} and a similar
behavior for the effective screening function was obtained.
Essentially, effective Coulomb interactions obtained in the present
work manifest the same asymptotic behavior as those obtained in Refs.
\cite{KI13,KI12} where the Thomas-Fermi equation with a fixed
boundary at infinity is solved.

We solved Eq.~(\ref{dirac}) in a screened Coulomb potential, with an
accurate numerical method presented in Refs. \cite{Sal91,Sal95}. The
method allows control of truncation errors of the solutions and the
only remaining uncertainties were due to unavoidable round-off
errors and due to the distortion of the potential introduced by the
interpolating spline. Detailed information about this method can be
found in our previous article \cite{Sab16}. Further, the integration
of the PSF values was performed accurately with Gauss-Legendre
quadrature in 32 points.

Because we compare our results with those of Gove and Martin
\cite{Gov71}  we mention selected features of their method of
calculation of the PSFs. They obtained the radial electron functions
($f$, $g$) as solutions of Dirac equations for a point-nucleus and
with an unscreened Coulomb spherical potential. The obtained
functions were approximate, expressed in terms of $\Gamma$
functions. According to their prescriptions \cite{Gov71}, the finite
nuclear size and screening effects were treated as corrections to
these approximate functions.

\begin{table}[h]
 \scriptsize   \caption{\scriptsize Comparison of PSF results calculated
 with our recipe (TW) and Gove and Martin recipe ($GM$), for $\beta^+$
decay of three virtual  nuclei.}\label{virtual_betaplus}

        \resizebox{8.5cm}{!}
{ \scriptsize
    \begin{tabular}{|cc|c|c|c|}
        \hline
        Z  &   A  &  $Q_{\beta^+} (MeV)$ & Log($F^{(GM)}_{\beta^+}$)\cite{Gov71} & Log($F^{(TW)}_{\beta^+}$) \\
        \hline
        10  & 20 &   0.05 &         -4.643 & -4.646\\
        &    &   0.50  &         -0.700  &  -0.702\\
        &    &    5.00   &         3.575  &  3.574\\
        \hline
        50  & 120&   0.05 &         -6.117 &  -6.195\\
        &    &   0.50  &         -1.152  &   -1.175\\
        &    &    5.00   &         3.319  &   3.311\\
        \hline
        90  & 230 &  0.05 &         -7.088 &   -7.272\\
        &     &  0.50  &         -1.330  &   -1.388 \\
        &     &   5.00   &         3.236  &    3.206\\
        \hline
    \end{tabular}
}
\end{table}

\begin{table}[h]

 \scriptsize \caption{\scriptsize Comparison of PSF results calculated with our
 recipe (TW) and ($GM$), for $\beta^-$ decay of
three virtual  nuclei.}\label{virtual_betaminus}

    \resizebox{8.5cm}{!}
    {
\scriptsize
        \begin{tabular}{|cc|c|c|c|}
            \hline
            Z  &   A  &  $Q_{\beta^-} (MeV)$ & Log($F^{(GM)}_{\beta^-}$)\cite{Gov71} & Log($F^{(TW)}_{\beta^-}$) \\
            \hline
            10  & 20 &   0.05 &         -3.755 & -3.793\\
            &    &   0.50  &         -0.366  &  -0.369\\
            &    &    5.00   &         3.776  &  3.774\\
            \hline
            50  & 120&   0.05 &         -2.776 &  -2.978\\
            &    &   0.50  &         0.389  &   0.307\\
            &    &    5.00   &         4.304  &   4.282\\
            \hline
            90  & 230 &  0.05 &         -1.857 &   -2.026\\
            &     &  0.50  &         1.269  &   1.106 \\
            &     &   5.00   &         4.929  &   4.904 \\
            \hline
        \end{tabular}
    }
\end{table}

In order to illustrate the differences that appears between our
calculation and the approximate method of $GM$
\cite{Gov71}, we compare the PSF results for three virtual cases as
discussed in Ref. \cite{Gov71}. As seen from
Table~\ref{virtual_betaplus} for $\beta^+$, and
Table~\ref{virtual_betaminus} for $\beta^-$, the differences between
the two sets of PSF values may not be so obvious in the decimal
logarithm scale, but in half-lives calculation where the absolute
PSF values are used,
the differences may be relevant as we will see in the next section.

\subsubsection{Phase space factors for electron capture (EC)}

Electron capture is a process which competes with positron decay. It
is an alternate decay mode for the $\beta^+$ unstable nuclei that do
not have enough energy to decay by positron emission. Considering
the fact that the electron capture from the M-, N- and higher shells
have negligible contributions in comparison with the K- and L- ones,
we can write the PSF expression of electron capture for an allowed
transition as

\begin{eqnarray}\label{eckl}
F_{EC}^{K,L_1} = {\pi\over 2} \left(q^2_K g^2_K B_K + q^2_{L_1} g^2_{L_1}
B_{L_1}\right).
\end{eqnarray}

For the $q_{K/L_1}$ quantities we used the expression

\begin{eqnarray}
q_{K/L_1}=W_{EC}-\epsilon_{K/L_1},
\end{eqnarray}
were, $W_{EC}$ is the Q value of the $\beta^+$ decay in $m_ec^2$
units, $\epsilon_i$ are the binding energies of the 1s$_{1/2}$ and
2s$_{1/2}$ electron orbitals of the parent nucleus, $g_i$ their
radial densities on the nuclear surface. $B_i\approx 1$ represent
the values of the exchange correction. In our method we consider these
exchange corrections to be unity, for the nuclei considered, the
estimated error in doing that being under 1\%. The relation
$W_0=W_{EC}-1$ holds.

The $g_{K/L_1}$ are the electron bound states, solutions of the
Dirac equation (\ref{dirac}), and correspond to the eigenvalues
$\epsilon_{n}$ ($n$ is the radial quantum number). The quantum
number $\kappa$ is related to the total angular momentum
$j_\kappa=\mid\kappa\mid-1/2$. These wave functions are normalized
such that

\begin{equation}
\int_0^\infty [g^2_{n,\kappa}(r)+f^2_{n,\kappa}(r)]dr=1.
\end{equation}

For the $EC$ processes, the potential used to obtain the electron
wave functions reads

\begin{equation}
rV_{EC}(Z,r)=rV(Z,r)\phi(r),
\end{equation}

\noindent and the charge number $Z=Z_0$ corresponds to the parent
nucleus. $V(Z,r)$ is negative. More details about the numerical
procedure can be found in \cite{Sab16}.

\section{\label{sec:level3}RESULTS AND DISCUSSION}

Half-lives were computed using Eq.~(\ref{hl}) and Eq.~(\ref{phl}).
The NMEs were calculated using Eq.~(\ref{ftp}) (for Fermi
transitions) and Eq.~(\ref{rtp}) (for GT transitions) within the
pn-QRPA formalism. For the PSF calculation we used two different
recipes. One is our newly calculation recipe \cite{Sab16} and the
other one is the conventional computation using the prescription of
$GM$ \cite{Gov71}. We again state that the same
set of NMEs were used in both types of half-life calculations.
For the $GM$ bound states, the Dirac equation is solved by assuming
a Hartree self consistent Coulomb field. The finite size effect
is introduced within a nuclear charge distribution of a
Woods-Saxon form. In the $GM$ model, the screening is
implicitly taken into account. In the calculations of this work, the Coulomb
field is obtained by considering a nuclear charge distribution
obtained within a shell model and a screening effect is introduced
by modifying the potential. Therefore, some differences in the
electron binding energies and in the wave functions arise between the
two recipes.
The electron energies and their radial densities on the nuclear surface
for $K$ and $L_1$ orbitals recalculated in this work with
recently improved numerical codes following the  recipe of Ref. \cite{Sab16}
are listed in Table \ref{tdens} and compared with those of \cite{Mar70}.

\begin{figure}[h]
    \includegraphics [width=3.5in]{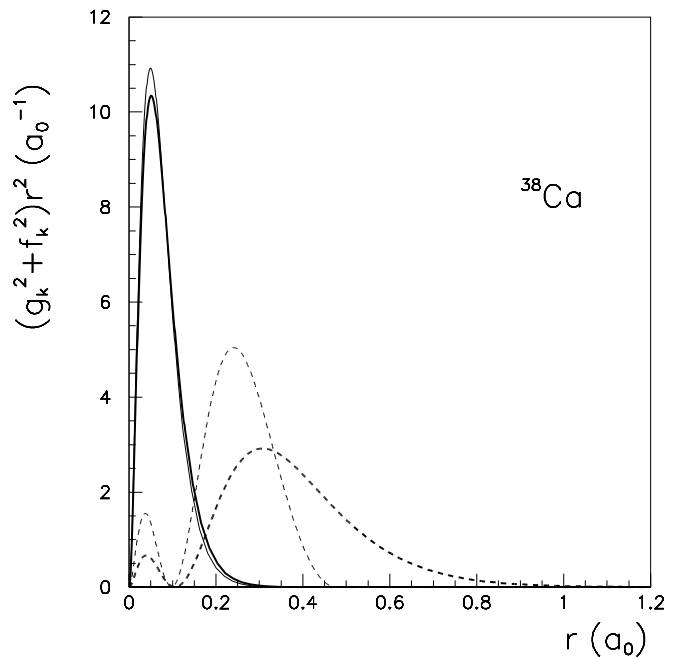}
    \centering \caption{\scriptsize
The radial distributions $(g_k^2+f_k^2)r^2$ are plotted for the
$K$ and $L_1$ shells as function of the distance $r$ in term
of atomic units $a_0$ for $^{38}$Ca. The thick lines
are obtained with the treatment used in this work. The thin curve is
used for the potential without screening and with a constant
proton density inside the nucleus. The full line is used for the
$K$ shell  while the dashed one corresponds to the $L_1$ orbital.
     \label{psir}
}
\end{figure}

The differences between the results obtained for a Coulomb field obtained from
a constant charge density inside a spherical nucleus and those
obtained with a field constructed with a more realistic charge density
corrected with a screening function can offer an estimation of the role
played by the ingredients of our model. The radial distributions of
the electron $(g_k^2+f_k^2)r^2$ are displayed in  Fig. \ref{psir}
for these two treatments in the case the parent nucleus $^{38}$Ca.
The radial distributions in the case of a pure Coulomb field, plotted with
thin lines,
are more confined in the vicinity of the nucleus, therefore the amplitudes
of the wave function are larger. These differences are translated in the values
$g^2_i$ and of the electron
binding energies $\epsilon_i$, as it can be seen in Table \ref{t38ca}.
Adding the screening effect in our calculations makes our results very close
to those of $GM$.

Table \ref{BPHL} presents a comparison between the measured and
calculated half-lives for  $\beta^{+}$/EC-decay of twenty medium and
heavy nuclei of interest. Entries in third column are calculated
using the pn-QRPA method for the NMEs, while the PSFs are calculated
by the by $(GM)$ method \cite{Gov71}. The
fourth column shows the calculated half-lives using our new recipe
of PSFs \cite{Sab16} and labeled $(TW)$ (this work). Most of the
nuclei shown in this table are the same as those presented in
Table~2 of Ref. \cite{Sab16}. All half-lives are given in units of
seconds. Q-values for the reaction were taken from \cite{Aud12}. It
is seen from Table~\ref{BPHL} that the newly calculated half-lives
are systematically larger than those computed using the PSFs of
$(GM)$ \cite{Gov71}. The last column displays the percentage
deviation (PD) of the two calculated half-lives. We calculated the
PD between the two computed half-lives using the formula

\begin{equation}
PD= \frac{T_{1/2}^{(TW)}-T_{1/2}^{(GM)}}{T_{1/2}^{(TW)}}\times 100
(\%) \label{PD}
\end{equation}

Table~\ref{BPHL} shows that the PD increases to a maximum value of
4.05 $\%$ for the $^{56}$Ni nucleus. The case of EC on $^{205}$Bi requires special mention. For this nucleus the pn-QRPA predicts couple of high excited transitions to the daughter nucleus and the available Q-value of these two transitions is lower than the binding energy of the K-shell electron. Using calculation, after the EC process of a K-shell electron, the neutrino may have a negative energy which is not physical. Accordingly we only calculate L-shell EC in this case.
In the calculated $\beta^{+}$/EC-decay half-lives,
the smallest difference was noted for $^{52}$Fe.
In Table~\ref{BPPHL} we show the state-by-state transitions for
two cases: $^{52}$Fe and $^{56}$Ni.
Shown are also the adopted NMEs using the pn-QRPA model,
the calculated PSFs (separately for both EC and $\beta^{+}$-decay
reactions), partial half-lives (PHL), Q-values and branching ratios
I$_{(\beta^{+}/EC)}$. The branching ratio '$I$' for each transition
was calculated using the formula

\begin{equation}
I = \frac{T_{1/2}}{t_{f}}\times 100 (\%),
\label{BR}
\end{equation}
where T$_{1/2}$ is the total $\beta$ decay half-life and t$_{f}$ is
the calculated partial half-life of the corresponding transition. For the
nucleus $^{56}$Ni we note calculation of much smaller PSF for
EC decay to daughter energies using our recipe. The
PSFs calculated from $(GM)$ recipe is on average within 3 \% smaller.
This in turn led to a 4 \% larger calculated half-life value
for $^{56}$Ni using our recipe. 

We would like to comment further on the entries in the first two columns of Table~\ref{BPPHL}. These entries are model dependent. The excited states in daughter nuclei (shown in the first column) and NMEs (presented in the second column) were calculated using the pn-QRPA model. The computed excited states satisfied the selection criteria for allowed transitions within the chosen model. A different nuclear model can change the entries in the first two columns and, as stated earlier, is not the focus of current study. Q-values are presented
in the third and the sixth columns of
Table~\ref{BPPHL}  using following relation

\begin{equation}
Q_{EC} = m_{p}-m_{d}-E_{f}, \label{QEC}
\end{equation}

and

\begin{equation}
Q_{BP} = m_{p}-m_{d}-E_{f}-2m_{e}c^{2}. \label{QBP}
\end{equation}
Here $m_{p}$ and
$m_{d}$ are masses of parent and daughter nuclei, respectively
whereas $E_{f}$ is the calculated energy levels in the daughter nucleus (model dependent).

Table~\ref{BMHL} shows the comparison of measured and calculated
half-lives for $\beta^-$ decay cases. The Q-values were taken from
\cite{Aud12}. The comparison between the two calculations is much
better for the $\beta^-$ decay half-lives than those correponding to
$\beta^+$.
Table~\ref{BMPHL} shows the state-by-state
calculation of PHL for $^{100}$Sr (largest PD=3.82 \%) and $^{152}$Nd
(PD=2.04 \%). The values of $Q_{\beta^-}$ (calculated as in
Eq.~(\ref{QEC})), PSF, NME and branching ratios I$_{(\beta^-)}$ are
also given in Table~\ref{BMPHL}. We note an overall  agreement
between our calculated PSF values and those using the $(GM)$ recipe
in  the case of of both analyzed nuclei.

\begin{figure*}[h]
    \includegraphics [width=6.5in]{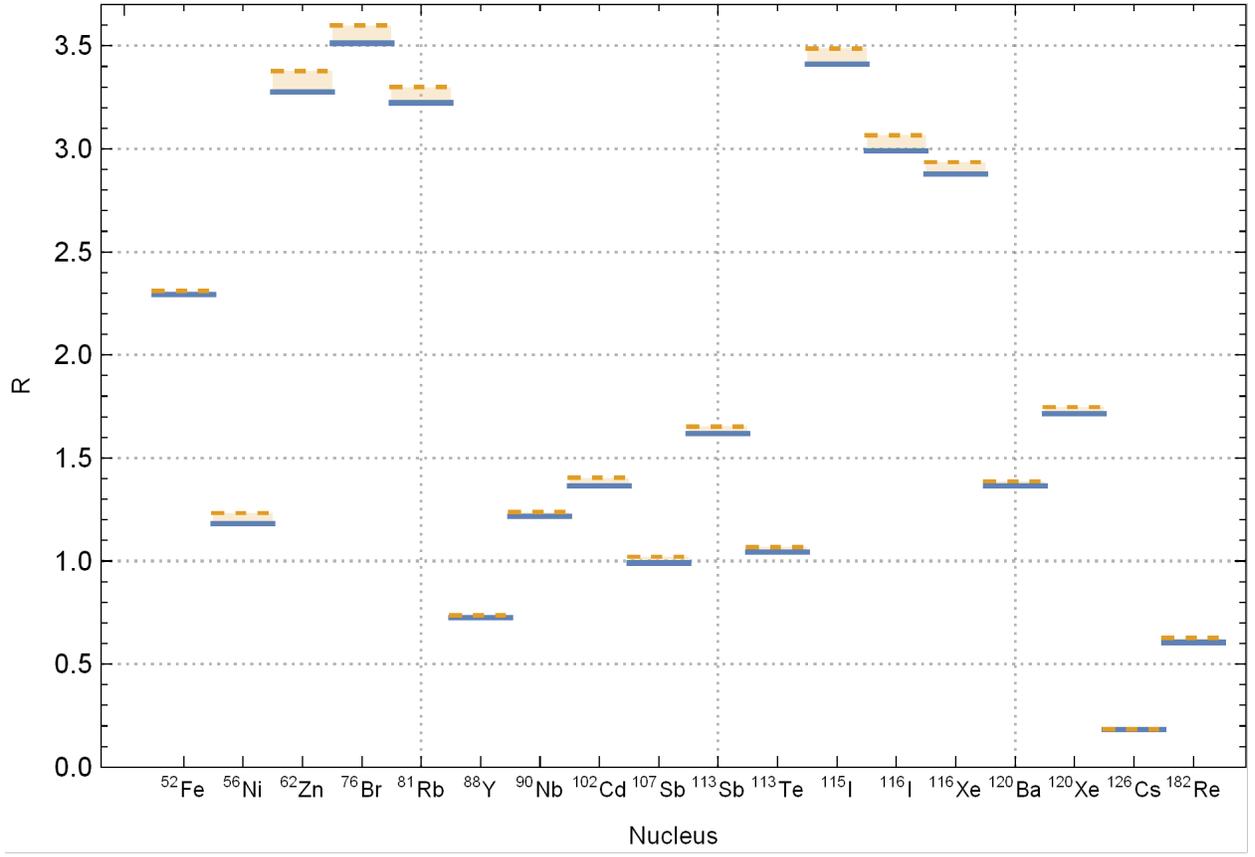}
    \caption{\scriptsize Ratios $R$ between experimental \cite{Aud12} and calculated half-lives undergoing  $\beta^+$ decay for selected cases. Full lines: theoretical half-lives calculated within the $(TW)$ recipe. Dotted lines: theoretical half-lives calculated with
the $(GM)$ recipe of Ref. \cite{Gov71}.} \label{BP_ratio}
\end{figure*}

\begin{figure*}[h]
    \includegraphics [width=6.5in]{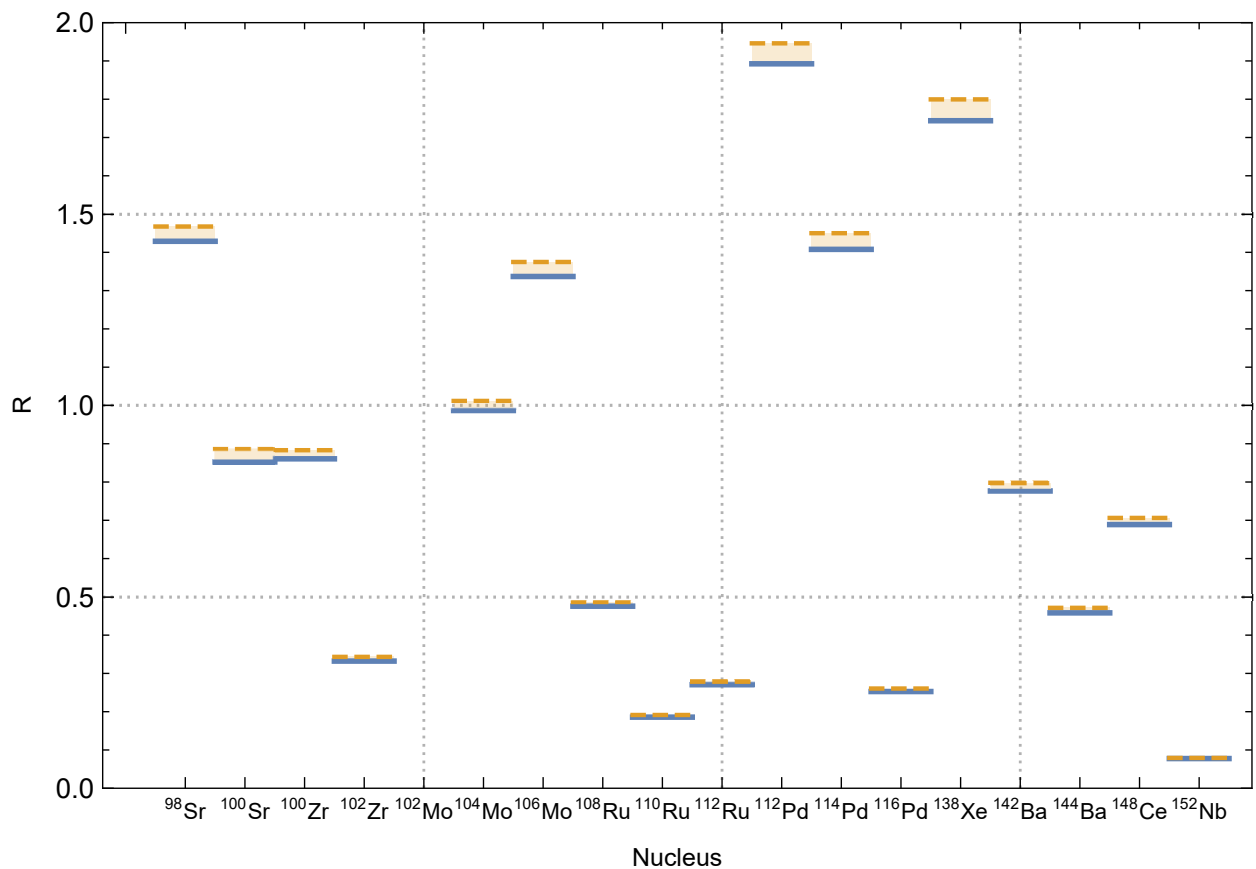}
    \centering \caption{\scriptsize Same as Fig.~\ref{BP_ratio} but for selected $\beta^{-}$ decay cases.} \label{BM_ratio}
\end{figure*}

The difference between the two calculated half-lives as well as
their mutual comparison with the experimental data, is done
intuitively in a graphical way in Fig.~\ref{BP_ratio} for few
selected  $\beta^+$ decay cases. We display the ratio between the
experimental half-lives and the theoretical ones,
$R=T_{1/2}^{(exp)}/T_{1/2}^{(X)}$, where $(X)$ stands for the
calculation recipe, $(GM)$ or $(TW)$. With solid lines the ratios
calculated with $(TW)$ recipe are represented, while with dotted
lines the conventional $(GM)$ computations are displayed.
 We note that systematically our half-lives are larger that the $GM$,
improving the agreement with the
measured data for most of the cases. From Fig.~\ref{BP_ratio} it can be remarked that the $(TW)$
ratios are in general closer to the value 1
than the $(GM)$ ones.
This effect is highlighted in
Fig.~\ref{BP_ratio} by the link between the dotted line and the
solid line. In Fig.~\ref{BM_ratio}, the ratios corresponding to
$\beta^-$ decays are displayed in the same manner as in
Fig.~\ref{BP_ratio}. It is noted that no appreciable improvement is
brought in calculation of $\beta^-$ decay half-lives except for a
few cases in which the experimental data are undervalued by calculations,
as also evident from Table~\ref{BMHL}. Overall, we note a
good agreement of the new theoretical half-lives with the
experimental ones.
It is again remarked that this
comparison could have improved further with a more reliable set of
NMEs or choice of better model parameters for the calculation of
NMEs (not the subject of current paper).

The differences between the $(GM)$ and the present results can be
explained by the use of a more rigorous approach in our case for the
free states in the PSF computation, but also are due to the differences
between our potential and the one used by $(GM)$. Regarding only the free states,
in the $(GM)$ method the screening
correction was introduced empirically, by modifying the solutions of
the Dirac equation with a function evaluated at the nuclear surface.
This function depends on the difference between the effective
potential and the point like nucleus  Coulomb interaction. On the
other hand, in our calculation the screening is introduced by
considering an effective Coulomb potential. The Dirac equation is
solved numerically for this effective Coulomb potential up to large
values of $r$, where the wave functions are well approximated by its
asymptotic form. Later, the wave functions are normalized by
comparing their value with the asymptotic forms. This normalization
determines the value of each wave function on the surface of the
nucleus. Further, in the $(GM)$ calculation, the nuclear finite size
of the nucleus is simulated by additional corrections to the Fermi
functions, while in our calculation, the effective Coulomb field is
built up from the proton density of the nucleus, as also mentioned
before. Regarding the bound states, the $(GM)$ method uses
tabulated values of the energies and of the radial densities that are
obtained by solving the Dirac equation within a more sophisticated
self-consistent Coulomb potential.

\section{\label{sec:level4}Summary and Conclusion}

The aim of present work was to investigate the effect of the
incorporation of new PSF values, computed with a more precise and
rigorous method, on the theoretical half-lives for $\beta^{\pm}$ and
EC decays of unstable nuclei. The newly calculated $\beta$-decay
half-lives were systematically larger than those given in the
previous calculations. The mean percentage deviation is larger for
the $\beta^-$ decay ($2.73\%$) as compared to the $\beta^+$/EC decay
rates ($2.35\%$). For the adopted set of NMEs, in general the
half-lives computed with newly PSFs are closer to the measured ones
than the half-lives calculated with PSFs with approximate method
(i.e., using approximate electron wave functions) \cite{Gov71} for
free states. Although the largest uncertainty in the computation of
$\beta$-decay half-lives comes from the NMEs, introduction of the
newly PSF values may improve the comparison with experiment and
should be taken into account for accurate predictions.

In near future we would be presenting the effect of calculation of
newly computed PSF on stellar beta decay rates and comment on its
astrophysical implications.\\

\textbf{Acknowledgments}\\

J.-U. Nabi would like to acknowledge the support of the Higher Education Commission Pakistan through project numbers 5557/KPK/NRPU/R$\&$D/HEC/2016,
9-5(Ph-1-MG-7)/PAK-TURK/R$\&$D/HEC/2017 and Pakistan Science
Foundation through project number PSF-TUBITAK/KP-GIKI (02). M. Ishfaq wishes to
acknowledge the support provided by Scientific and Technological
Research Council of Turkey (TUBITAK), Department of Science
Fellowships and Grant Programs (BIDEB) 2216 Research Fellowship
Program For International Researchers (21514107-115.02-124287).

S. Stoica, M. Mirea and O. Ni\c{t}escu would like to acknowledge the
support of the Romanian Ministry of Research and Innovation,
CNCS-UEFISCDI, through projects PN-III-P4-ID-PCE-2016-0078 and PN-5N/2018.


\onecolumngrid
 \clearpage
 \centering
 \begin{table}
        \scriptsize \caption{\scriptsize The first four columns
give the electron binding energies $\epsilon_i$ for the shells $K$ and $L_1$
                and their radial wave function densities $g_i$
                at the nuclear surface calculated in this work for different nuclei. The last two columns give the values of Ref. \cite{Mar70}.
        }
        \label{tdens}
        \begin{tabular}{|c|c|c|c|c|c|c|} \hline
                Nucleus & $\epsilon_K$ (keV) & $\epsilon_{L_1}$ (keV)& $g_K^2$ ($[\hbar/mc]^{-3}$)& $g_{L_1}^2/g^2_{K}$ & $g_K^2$ ($[\hbar/mc]^{-3}$) \cite{Mar70}& $g_{L_1}^2/g^2_{K}$ \cite{Mar70} \\ \hline
                $^{52}$Fe  & 6.63991 & 0.7451126 & 0.0327063 & 0.0898416   & 0.0328     & 0.0950        \\
                $^{56}$Ni  & 7.95277 & 0.9181003 & 0.0411267 & 0.0943036   & 0.0423     & 0.0974        \\
                $^{62}$Zn  & 9.20973 & 1.1092315 & 0.0530401 & 0.0953901   & 0.0538     & 0.0995        \\
                $^{76}$Br  & 13.0121 & 1.6760343 & 0.0917936 & 0.1019465   & 0.0935     & 0.1035        \\
                $^{81}$Rb  & 14.6718 & 1.9336296 & 0.1136202 & 0.1042520   & 0.1149     & 0.1063        \\
                $^{88}$Y   & 16.4688 & 2.0221700 & 0.1389702 & 0.1272602   & 0.1402     & 0.1080        \\
                $^{90}$Nb  & 18.3994 & 2.4889101 & 0.1686632 & 0.1121010   & 0.170      & 0.1098        \\
                $^{102}$Cd & 26.1177 & 3.9008765 & 0.3182812 & 0.1109908   & 0.319      & 0.1159         \\
                $^{105}$Ag & 24.95904 & 3.558919 & 0.2900978 & 0.1193451   & 0.293      & 0.1150         \\
                $^{107}$Sb & 29.99173 & 4.140248 & 0.4109690& 0.1416423    & 0.413      & 0.1187        \\
                $^{113}$Sb & 29.99173 & 4.140248 & 0.4101592& 0.1416413    & 0.413      & 0.1187        \\
                $^{113}$Te & 31.18294 & 4.70109 & 0.4488353 & 0.12019908   & 0.449      & 0.1196        \\
                $^{115}$I  & 32.50419 & 4.937340 & 0.4894928& 0.1210013    & 0.488      & 0.1205        \\
                $^{116}$I  & 32.50419 & 4.937340 & 0.4893257& 0.1210012    & 0.488      & 0.1205        \\
                $^{116}$Xe & 33.95443 & 5.055330 & 0.5289785& 0.1300301    & 0.529      & 0.1215        \\
                $^{120}$Ba & 36.81175 & 4.975840 & 0.6244052 & 0.1662902   & 0.623      & 0.1234        \\
                $^{120}$Xe & 33.95440 & 5.055328 & 0.5280176& 0.1300293    & 0.529      & 0.1215        \\
                $^{126}$Cs & 35.30411 & 5.151703 & 0.5764054& 0.1394027    & 0.574      & 0.1224        \\
                $^{182}$Re & 71.29588 & 12.23093 & 2.7488862& 0.1490819    & 2.69       & 0.1448        \\
                $^{205}$Bi & 90.39904 & 16.18943 & 5.0324494& 0.1587003    & 4.88       & 0.1561        \\
                \hline
        \end{tabular}
\end{table}

\begin{table}
\caption{The results $GM$ obtained in Ref. \cite{Mar70} are compared with
the results $TW$ of this work and those obtained without a realistic charge density
and without screening correction in the case of $^{38}$Ca.
The $K$ and $L_1$  shells are illustrated. The binding
energies are denoted $\epsilon_i$ and the radial densities
on the surface of the nucleus are denoted $g_i^2$.}
\label{t38ca}
\begin{tabular}{|c|c|c|c|c|}
\hline
Method & $\epsilon_K$ (keV) & $\epsilon_{L1}$ (keV) & $g_{K}^2$ ($[\hbar/mc]^{-3}$) & $g_{L1}^2/g_{K}^2$ \\ \hline
Ref. \cite{Mar70} & 3.60740 & 0.37710 & 0.01367 & 0.08620 \\
$TW$   & 3.762904& 0.357179 & 0.01308474 & 0.0825267 \\
No screening & 5.454032 & 1.198887 & 0.01434885 & 0.1877057 \\ \hline
\end{tabular}
\end{table}

\begin{table}
   \scriptsize \caption{\scriptsize Comparison of measured, calculated half-lives and
        percentage deviation (PD) for $\beta^{+}$/EC-decay of selected
        nuclei. For the case of $^{205}$Bi we calculate only L-shell EC.}\label{BPHL}

    \begin{tabular}{|c|c|c|c|c|c|}
        \hline
        Nucleus & $T_{1/2}^{(EXP)}$ (s) \cite{Aud12} & $T_{1/2}^{(GM)}$ (s) \cite{Gov71}& $T_{1/2}^{(TW)}$ (s)  & PD (\%) \\
        \hline $^{52}$Fe     & 2.98E+04 &1.29E+04         &1.30E+04  & 0.77 \\
        $^{56}$Ni            & 5.25E+05 &4.26E+05         &4.44E+05  & 4.05 \\
        $^{62}$Zn            & 3.31E+04 &9.80E+03         &1.01E+04  & 2.97 \\
        $^{76}$Br            & 5.83E+04 &1.62E+04         &1.66E+04  & 2.41 \\
        $^{81}$Rb            & 1.65E+04 &5.00E+03         &5.12E+03  & 2.34 \\
        $^{88}$Y             & 9.21E+06 &1.25E+07         &1.27E+07  & 1.57 \\
        $^{90}$Nb            & 5.26E+04 &4.25E+04         &4.32E+04  & 1.62 \\
        $^{102}$Cd           & 3.30E+02 &2.35E+02         &2.42E+02  & 2.89 \\
        $^{105}$Ag           & 3.57E+06 &2.45E+04         &2.52E+04  & 2.78 \\
        $^{107}$Sb           & 4.00E+00 &3.92E+00         &4.04E+00  & 2.97\\
        $^{113}$Sb           & 4.00E+02 &2.42E+02         &2.47E+02  & 2.02\\
        $^{113}$Te           & 1.02E+02 &9.55E+01         &9.77E+01  & 2.25\\
        $^{115}$I            & 3.48E+02 &9.98E+01         &1.02E+02  & 2.16\\
        $^{116}$I            & 2.91E+00 &9.49E-01         &9.73E-01  & 2.47\\
        $^{116}$Xe           & 5.90E+01 &2.01E+01         &2.05E+01  & 1.95\\
        $^{120}$Ba           & 2.40E+01 &1.73E+01         &1.76E+01  & 1.70\\
        $^{120}$Xe           & 2.76E+03 &1.58E+03         &1.61E+03  & 1.86 \\
        $^{126}$Cs           & 9.84E+01 &5.35E+02         &5.42E+02  & 1.29 \\
        $^{182}$Re           & 2.30E+05 &3.67E+05         &3.80E+05  & 3.42 \\
        $^{205}$Bi           & 1.32E+06 &1.47E+06         &1.52E+06  & 3.46 \\
        \hline
    \end{tabular}
\end{table}


\centering

\clearpage \centering

\clearpage \scriptsize
\begin{table}
    \scriptsize\caption{\scriptsize State-by-state comparison of calculated PSF (for  $\beta^{+}$/EC-decay) using recipe of \cite{Gov71} and current prescription (TW).
    Shown also are the daughter energy levels, nuclear matrix elements NME,  $Q$ values, partial half-lives
    (PHL) for  $\beta^{+}$/EC-decay and branching ratio I$_{(\beta^{+}/EC)}$ of the selected nuclei.}\label{BPPHL}

    \begin{tabular}{|c|c|c|c|c|c|c|c|c|c|c|c|c|c|}
        \hline$^{52}$Fe\\
     \hline E$_{x}$(MeV) & NME & $Q_{EC}$ (MeV)& $F^{(GM)}_{EC}$\cite{Gov71}    & $F^{(TW)}_{EC}$ & $Q_{\beta^+}$ (MeV) & $F^{(GM)}_{\beta^+}$\cite{Gov71} & $F^{(TW)}_{\beta^+}$ & PHL$^{(GM)}$\cite{Gov71} & PHL$^{(TW)}$ & I$^{(GM)}_{(\beta^{+}/EC)}$ \cite{Gov71} & I$^{(TW)}_{(\beta^{+}/EC)}$\\
    \hline              0.000      & 0.02768& 2.3733& 1.22206 &1.20150 &   1.3512            &  8.41032  &  8.31627   & 1.50147E+04 & 1.51955E+04 & 85.701&85.713\\
                        0.004      & 0.00170& 2.3693& 1.21794 &1.19744 &   1.3472            &  8.30402  &  8.21074   & 2.46699E+05& 2.49683E+05 & 5.2160&5.2160\\
                        0.196      & 0.00325& 2.1773& 1.02771 &1.01077 &   1.1552            &  4.29193  &  4.24688   & 2.31350E+05& 2.34078E+05 & 5.5620&5.5640\\
                        0.291      & 0.00134& 2.0823& 0.94007 &0.92425 &   1.0602            &  2.98166  &  2.94390   & 7.60563E+05& 7.71098E+05 & 1.6920&1.6890\\
                        0.720      & 0.00253& 1.6533& 0.59198 &0.58175 &   0.6312            &  0.33565  &  0.32942   & 1.70770E+06& 1.73853E+06 & 0.7540&0.7490\\
                        0.939      & 0.00087& 1.4343& 0.44543 &0.43734 &   0.4122            &  5.69E-02  &  5.53E-02  & 9.20851E+06& 9.39057E+06 & 0.1400 &0.1390\\
                        1.011      & 0.00350& 1.3623& 0.40124 &0.39435 &   0.3402            &  2.53E-02  &  2.47E-02  & 2.67938E+06& 2.72749E+06 & 0.4800&0.4780\\
                        1.362      & 0.00255& 1.0113& 0.22076 &0.21663 &  -0.0107            &  -  &  -   & 7.10887E+06& 7.24422E+06 & 0.1810 &0.1800\\
                        1.467      & 0.00052& 0.9063& 0.17691 &0.17374 &  -0.1157            &  -  &  -   & 4.36772E+07& 4.44742E+07 & 0.0290 &0.0290\\
                        1.685      & 0.00017& 0.6883& 1.01706E-01 &9.97773E-02 & -0.3371         &  -  &  -   & 2.35223E+08& 2.39771E+08 & 0.0050 &0.0050\\
                        1.754      & 0.00009& 0.6193& 8.22094E-02 &8.06130E-02 & -0.4027         &  -  &  -   & 5.31896E+08& 5.42429E+08 & 0.0020 &0.0020\\
                        1.821      & 0.00972& 0.5523& 6.54251E-02 &6.39581E-02 & -0.4697         &  -  &  -   & 6.29314E+06& 6.43748E+06 & 0.2040 &0.2020\\
                        2.119      & 0.00282& 0.2543& 1.35835E-02 &1.32041E-02 & -0.7677         &  -  &  -   & 1.04562E+08& 1.07568E+08 & 0.0120 &0.0120\\
                        2.143      & 0.00572& 0.2303& 1.10469E-02 &1.07735E-02 & -0.7917         &  -  &  -   & 6.33834E+07& 6.49917E+07 & 0.0200 &0.0200\\
    \hline$^{56}$Ni\\ \hline 1.196      & 0.00038 & 0.9357 & 0.24313       &  0.23335         &  -0.0862    &  -  &  -   & 4.31594E+07&4.49676E+07 &0.986 &0.988\\
                        1.247      & 0.00019 & 0.8847 & 0.21729       &  0.20842         &  -0.1372    &  -  &  -   & 9.85234E+07&1.02717E+08 &0.432 &0.432\\
                        1.252      & 0.00034 & 0.8797 & 0.21476       &  0.20605         &  -0.1422    &  -  &  -   & 5.55046E+07&5.78528E+07 &0.767 &0.768\\
                        1.288      & 0.00009 & 0.8437 & 0.19721       &  0.18939         &  -0.1782    &  -  &  -   & 2.13703E+08&2.22525E+08 &0.199 &0.200\\
                        1.289      & 0.00009 & 0.8427 & 0.19676       &  0.18894         &  -0.1792    &  -  &  -   & 2.31551E+08&2.41134E+08 &0.184 &0.184\\
                        1.299      & 0.00008 & 0.8327 & 0.19227       &  0.18444         &  -0.1892    &  -  &  -   & 2.69031E+08&2.80450E+08 &0.158 &0.158\\
                        1.309      & 0.00001 & 0.8227 & 0.18754       &  0.18000         &  -0.1992    &  -  &  -   & 1.73974E+09&1.81258E+09 &0.024 &0.024\\
                        1.313      & 0.00002 & 0.8187 & 0.18593       &  0.17824         &  -0.2032    &  -  &  -   & 1.22606E+09&1.27894E+09 &0.035 &0.035\\
                        1.318      & 0.00001 & 0.8137 & 0.18344       &  0.17605         &  -0.2082    &  -  &  -   & 1.76475E+09&1.83885E+09 &0.024 &0.024\\
                        1.363      & 0.00000 & 0.7687 & 0.16374       &  0.15695         &  -0.2532    &  -  &  -   & 4.34936E+10&4.53747E+10 &0.001 &0.001\\
                        1.363      & 0.00000 & 0.7687 & 0.16372       &  0.15695         &  -0.2532    &  -  &  -   & 2.36127E+10&2.46320E+10 &0.002 &0.002\\
                        1.373      & 0.00000 & 0.7587 & 0.15927       &  0.15285         &  -0.2632    &  -  &  -   & 7.54380E+10&7.86020E+10 &0.001 &0.001\\
                        1.471      & 0.00000 & 0.6607 & 0.12034       &  0.11558     &  -0.3612    &  -  &  -   & 1.51481E+10&1.57716E+10 &0.003 &0.003\\
                        1.482      & 0.00001 & 0.6497 & 0.11645       &  0.11172     &  -0.3722    &  -  &  -   & 3.31503E+09&3.45531E+09 &0.013 &0.013\\
                        1.503      & 0.00001 & 0.6287 & 0.10904       &  0.10454     &  -0.3932    &  -  &  -   & 3.72258E+09&3.88306E+09 &0.011 &0.011\\
                        1.711      & 0.07180 & 0.4207 & 4.82788E-02   &  4.62700E-02     &  -0.6012    &  -  &  -   & 1.15477E+06&1.20491E+06 &36.85 &36.85\\
                        1.742      & 0.13730 & 0.3897 & 4.13076E-02   &  3.95910E-02     &  -0.6322    &  -  &  -   & 7.05812E+05&7.36415E+05 &60.30 &60.30\\
                        \hline
    \end{tabular}
\end{table}

\centering


\clearpage \centering

\clearpage \scriptsize

\begin{table}
    \scriptsize \centering \caption{ \scriptsize Same as Table.~\ref{BPHL} but for
        $\beta^{-}$-decaying nuclei.}\label{BMHL}

    \begin{tabular}{|c|c|c|c|c|c|}
        \hline
        Nucleus &  $T_{1/2}^{EXP}$ (s) \cite{Aud12} &  $T_{1/2}^{(GM)}$ (s) \cite{Gov71}& $T_{1/2}^{(TW)}$ (s) & PD (\%)  \\
        \hline $^{98}$Sr  &6.53E-01 & 4.45E-01         & 4.57E-01 & 2.63        \\
        $^{100}$Sr        &2.02E-01 & 2.28E-01         & 2.37E-01 & 3.82    \\
        $^{100}$Zr        &7.10E+00 & 8.04E+00         & 8.25E+00 & 2.55  \\
        $^{102}$Zr        &2.90E+00 & 8.45E+00         & 8.73E+00 & 3.21  \\
        $^{102}$Mo        &6.78E+02 & 1.90E+02         & 1.94E+02 & 2.06  \\
        $^{104}$Mo        &6.00E+00 & 5.93E+00         & 6.08E+00 & 2.47 \\
        $^{106}$Mo        &8.73E+00 & 6.35E+00         & 6.53E+00 & 2.76 \\
        $^{108}$Ru        &2.73E+02 & 5.61E+02         & 5.74E+02 & 2.26 \\
        $^{110}$Ru        &1.20E+01 & 6.27E+01         & 6.46E+01 & 2.94 \\
        $^{112}$Ru        &1.75E+00 & 6.27E+00         & 6.47E+00 & 3.09 \\
        $^{112}$Pd        &7.57E+04 & 3.89E+04         & 4.00E+04 & 2.75 \\
        $^{114}$Pd        &1.45E+02 & 1.00E+02         & 1.03E+02 & 2.91 \\
        $^{116}$Pd        &1.18E+01 & 4.53E+01         & 4.67E+01 & 3.00 \\
        $^{138}$Xe        &8.44E+02 & 4.69E+02         & 4.84E+02 & 3.10 \\
        $^{140}$Xe        &1.36E+01 & 1.36E+00         & 1.40E+00 & 2.86 \\
        $^{142}$Ba        &6.36E+02 & 7.97E+02         & 8.19E+02 & 2.69\\
        $^{144}$Ba        &1.15E+01 & 2.44E+01         & 2.51E+01 & 2.79 \\
        $^{146}$Ce        &8.11E+02 & 4.74E+01         & 4.85E+01 & 2.27 \\
        $^{148}$Ce        &5.60E+01 & 7.93E+01         & 8.13E+01 & 2.46 \\
        $^{152}$Nd        &6.84E+02 & 8.64E+03         & 8.82E+03 & 2.04 \\
        \hline
    \end{tabular}
\end{table}


\clearpage \centering

\clearpage \scriptsize
\begin{table}
\scriptsize
    \caption{ \scriptsize State-by-state comparison of calculated PSF (for $\beta$-decay) using recipe of \cite{Gov71} and current prescription (TW).
    Shown also are the daughter energy levels, nuclear matrix elements NME,  $Q$ values, partial half-lives
    (PHL)  and branching ratio I$_{(\beta^-)}$ for $\beta^{-}$-decay of the selected nuclei.}\label{BMPHL}

    \begin{tabular}{|c|c|c|c|c|c|c|c|c|c|}
        \hline$^{100}$Sr\\
        \hline E$_{x}$(MeV) & NME & $Q_{\beta^-}$ (MeV) & $F^{(GM)}_{\beta^{-}}$\cite{Gov71} & $F^{(TW)}_{\beta^{-}}$ & PHL$^{(GM)}$\cite{Gov71} & PHL$^{(TW)}$ &  I$^{(GM)}_{(\beta^-)}$ \cite{Gov71} & I$^{(TW)}_{(\beta^-)}$\\
        \hline         0.13300 & 0.01473    &7.50300  &  86906.3& 72966.7 & 3.12798E+00& 3.72554E+00& 7.2730&6.3720\\
                       0.35700 & 0.00115    &7.14579  &  69557.1& 64334.6 & 4.99967E+01& 5.40553E+01& 0.4550&0.4390\\
                       0.86400 & 0.00510    &6.63935  &  49775.9& 47317.7 & 1.57827E+01& 1.66026E+01& 1.4410&1.4300\\
                       1.00300 & 0.00322    &6.50023  &  45212.0& 43184.3 & 2.75281E+01& 2.88207E+01& 0.8260&0.8240\\
                       1.06300 & 0.00429    &6.44024  &  43349.3& 41485.6 & 2.15140E+01& 2.24805E+01& 1.0570&1.0560\\
                       1.10400 & 0.00278    &6.39893  &  42102.4& 40346.8 & 3.42401E+01& 3.57299E+01& 0.6640&0.6640\\
                       1.35000 & 0.26502    &6.15278  &  35247.6& 33999.4 & 4.28548E-01& 4.44280E-01& 53.084&53.434\\
                       1.41900 & 0.01521    &6.08400  &  33499.6& 32359.0 & 7.85740E+00& 8.13437E+00& 2.8950&2.9180\\
                       1.62000 & 0.00002    &5.88283  &  28778.0& 27883.9 & 7.89102E+03& 8.14407E+03& 0.0030&0.0030\\
                       1.64200 & 0.00000    &5.86112  &  28301.8& 27429.7 & 2.05672E+05& 2.12210E+05& 0.0000&0.0000\\
                       1.66900 & 0.00080    &5.83372  &  27709.5& 26864.9 & 1.81631E+02& 1.87341E+02& 0.1250&0.1270\\
                       2.03600 & 0.03084    &5.46720  &  20688.6& 20109.3 & 6.27339E+00& 6.45412E+00& 3.6260&3.6780\\
                       2.19100 & 0.00096    &5.31207  &  18180.4& 17678.5 & 2.30158E+02& 2.36693E+02& 0.0990&0.1000\\
                       2.21100 & 0.00002    &5.29152  &  17867.1& 17374.4 & 1.26201E+04& 1.29779E+04& 0.0020&0.0020\\
                       2.21700 & 0.00126    &5.28623  &  17787.1& 17296.8 & 1.79065E+02& 1.84140E+02& 0.1270&0.1290\\
                       2.33500 & 0.01496    &5.16812  &  16074.2& 15633.6 & 1.66481E+01& 1.71173E+01& 1.3660&1.3870\\
                       2.40500 & 0.00704    &5.09834  &  15125.2& 14712.6 & 3.75885E+01& 3.86428E+01& 0.6050&0.6140\\
                       2.48600 & 0.05959    &5.01712  &  14077.3& 13695.1 & 4.77186E+00& 4.90505E+00& 4.7670&4.8400\\
                       2.55600 & 0.00015    &4.94728  &  13222.8& 12865.6 & 2.04511E+03& 2.10190E+03& 0.0110&0.0110\\
                       2.59700 & 0.00001    &4.90586  &  12735.9& 12392.7 & 3.58668E+04& 3.68600E+04& 0.0010&0.0010\\
                       2.71000 & 0.00051    &4.79275  &  11477.9& 11170.8 & 6.88307E+02& 7.07229E+02& 0.0330&0.0340\\
                       2.73300 & 0.04131    &4.76967  &  11233.8& 10933.7 & 8.62714E+00& 8.86390E+00& 2.6370&2.6780\\
                       2.92100 & 0.00092    &4.58202  &  9397.47& 9150.19 & 4.64906E+02& 4.77470E+02& 0.0490&0.0500\\
                       3.04900 & 0.00007    &4.45378  &  8285.68& 8069.38 & 7.37202E+03& 7.56963E+03& 0.0030&0.0030\\
                       3.19200 & 0.00721    &4.31146  &  7176.38& 6990.87 & 7.73167E+01& 7.93683E+01& 0.2940&0.2990\\
                       3.29000 & 0.00845    &4.21307  &  6480.88& 6314.67 & 7.30755E+01& 7.49989E+01& 0.3110&0.3170\\
                       3.29600 & 0.01583    &4.20677  &  6438.24& 6273.21 & 3.92753E+01& 4.03085E+01& 0.5790&0.5890\\
                       3.36400 & 0.00058    &4.13861  &  5990.84& 5838.16 & 1.15661E+03& 1.18685E+03& 0.0200&0.0200\\
                       3.43900 & 0.00029    &4.06433  &  5531.71& 5391.47 & 2.52582E+03& 2.59152E+03& 0.0090&0.0090\\
                       3.45000 & 0.01045    &4.05281  &  5463.09& 5324.69 & 7.01033E+01& 7.19255E+01& 0.3250&0.3300\\
                       3.47100 & 0.00612    &4.03206  &  5341.13& 5205.96 & 1.22440E+02& 1.25619E+02& 0.1860&0.1890\\
                       3.49600 & 0.00482    &4.00734  &  5198.67& 5067.29 & 1.59594E+02& 1.63732E+02& 0.1430&0.1450\\
                       3.56900 & 0.00134    &3.93407  &  4794.07& 4673.52 & 6.24537E+02& 6.40647E+02& 0.0360&0.0370\\
                       3.68200 & 0.11627    &3.82086  &  4218.21& 4112.92 & 8.16217E+00& 8.37111E+00& 2.7870&2.8360\\
                       3.88800 & 0.00006    &3.61534  &  3313.42& 3232.01 & 2.10528E+04& 2.15831E+04& 0.0010&0.0010\\
                       4.05100 & 0.01897    &3.45234  &  2711.28& 2645.88 & 7.78203E+01& 7.97437E+01& 0.2920&0.2980\\
                       4.08800 & 0.37148    &3.41515  &  2586.91& 2524.78 & 4.16561E+00& 4.26810E+00& 5.4610&5.5620\\
                       4.12800 & 0.20740    &3.37510  &  2458.04& 2399.32 & 7.85229E+00& 8.04448E+00& 2.8970&2.9510\\
                       4.16900 & 0.00354    &3.33446  &  2332.50& 2277.04 & 4.84389E+02& 4.96188E+02& 0.0470&0.0480\\
                       4.22200 & 0.00034    &3.28135  &  2176.15& 2124.70 & 5.47159E+03& 5.60410E+03& 0.0040&0.0040\\
                       4.30800 & 0.00556    &3.19476  &  1939.11& 1893.63 & 3.71131E+02& 3.80044E+02& 0.0610&0.0620\\
                       4.31100 & 0.00722    &3.19210  &  1932.17& 1886.87 & 2.87049E+02& 2.93941E+02& 0.0790&0.0810\\
                       4.40800 & 0.00101    &3.09502  &  1691.86& 1652.46 & 2.33941E+03& 2.39520E+03& 0.010&0.010\\
                       4.43400 & 0.00003    &3.06860  &  1630.72& 1592.78 & 8.24004E+04& 8.43631E+04& 0.000&0.000\\
                       4.53600 & 0.00007    &2.96655  &  1410.81& 1378.03 & 4.15176E+04& 4.25053E+04& 0.001&0.001\\
                       4.56000 & 0.00318    &2.94261  &  1362.77& 1331.13 & 9.23619E+02& 9.45568E+02& 0.025&0.025\\
                       4.62800 & 0.00284    &2.87494  &  1233.96& 1205.44 & 1.14035E+03& 1.16733E+03& 0.020&0.020\\
                       4.63100 & 0.03599    &2.87213  &  1228.83& 1200.43 & 9.05090E+01& 9.26499E+01& 0.251&0.256\\
                       4.68800 & 0.00938    &2.81463  &  1127.45& 1101.55 & 3.78490E+02& 3.87390E+02& 0.060&0.061\\
                       4.70300 & 0.01517    &2.79959  &  1102.05& 1076.78 & 2.39492E+02& 2.45112E+02& 0.095&0.097\\
                       4.78400 & 0.00317    &2.71946  &  974.255& 952.119 & 1.29597E+03& 1.32610E+03& 0.018&0.018\\
                       4.86000 & 0.33212    &2.64278  &  863.180& 843.718 & 1.39636E+01& 1.42857E+01& 1.629&1.662\\
                       4.87900 & 0.15890    &2.62366  &  837.107& 818.264 & 3.00956E+01& 3.07887E+01& 0.756&0.771\\
                       4.98800 & 0.04141    &2.51538  &  700.916& 685.342 & 1.37918E+02& 1.41052E+02& 0.165&0.168\\
                       4.99400 & 0.00028    &2.50876  &  693.199& 677.809 & 2.02785E+04& 2.07390E+04& 0.001&0.001\\
                       5.02400 & 0.00000    &2.47918  &  659.521& 644.946 & 2.18678E+06& 2.23620E+06& 0.000&0.000\\
                       5.02400 & 0.00000    &2.47907  &  659.398& 644.827 & 3.04674E+06& 3.11559E+06& 0.000&0.000\\
                       5.07300 & 0.00072    &2.42998  &  606.361& 593.032 & 9.17207E+03& 9.37823E+03& 0.002&0.003\\
                       5.07800 & 0.06105    &2.41548  &  601.620& 588.399 & 1.08997E+02& 1.11446E+02& 0.209&0.213\\
                       5.08800 & 0.00814    &2.41543  &  591.359& 578.371 & 8.31303E+02& 8.49970E+02& 0.027&0.028\\
                       5.14500 & 0.00755    &2.35801  &  534.741& 523.045 & 9.90972E+02& 1.01313E+03& 0.023&0.023\\
                       5.18900 & 0.00337    &2.31408  &  494.418& 483.653 & 2.40044E+03& 2.45387E+03& 0.009&0.010\\
                       5.22000 & 0.02056    &2.28304  &  467.396& 457.267 & 4.16606E+02& 4.25834E+02& 0.055&0.056\\
                       5.24100 & 0.01309    &2.26217  &  449.896& 440.172 & 6.79596E+02& 6.94608E+02& 0.033&0.034\\
                       5.31000 & 0.00883    &2.19284  &  395.410& 386.919 & 1.14639E+03& 1.17155E+03& 0.020&0.020\\
                       5.32100 & 0.00091    &2.18167  &  387.143& 378.840 & 1.14066E+04& 1.16566E+04& 0.002&0.002\\
                       5.33200 & 0.00363    &2.17111  &  379.456& 371.325 & 2.90418E+03& 2.96777E+03& 0.008&0.008\\
                       5.38200 & 0.19989    &2.12061  &  344.315& 336.988 & 5.81626E+01& 5.94271E+01& 0.391&0.399\\
                       5.42200 & 0.29697    &2.08095  &  318.556& 311.801 & 4.23152E+01& 4.32320E+01& 0.538&0.549\\
                       5.43600 & 0.00100    &2.06672  &  309.690& 303.131 & 1.28840E+04& 1.31628E+04& 0.002&0.002\\
                       5.44100 & 0.01497    &2.06191  &  306.737& 300.243 & 8.71971E+02& 8.90832E+02& 0.026&0.027\\
                       5.47800 & 0.02798    &2.02462  &  284.592& 278.578 & 5.02652E+02& 5.13504E+02& 0.045&0.046\\
                       5.51600 & 0.00603    &1.98731  &  263.713& 258.144 & 2.51817E+03& 2.57250E+03& 0.009&0.009\\
                       5.58100 & 0.00030    &1.92248  &  230.309& 225.470 & 5.86286E+04& 5.98869E+04& 0.000&0.000\\
                       5.58200 & 0.00028    &1.92101  &  229.593& 224.770 & 6.27505E+04& 6.40971E+04& 0.000&0.000\\
                       5.68100 & 0.00107    &1.82168  &  185.048& 181.209 & 2.02077E+04& 2.06358E+04& 0.001&0.001\\
                       5.68800 & 0.00137    &1.81513  &  182.369& 178.592 & 1.60203E+04& 1.63591E+04& 0.001&0.001\\
                       5.77800 & 0.00050    &1.72493  &  148.469& 145.430 & 5.40763E+04& 5.52065E+04& 0.000&0.000\\
                       5.83100 & 0.34429    &1.67201  &  131.012& 128.381 & 8.87476E+01& 9.05664E+01& 0.256&0.262\\
                       6.01400 & 0.87770    &1.48870  &  82.5496& 80.9479 & 5.52508E+01& 5.63440E+01& 0.412&0.421\\
                       6.03600 & 0.00558    &1.46719  &  77.9438& 76.4297 & 9.19758E+03& 9.37979E+03& 0.002&0.003\\
                       6.08500 & 0.00058    &1.41812  &  68.1847& 66.8508 & 1.01248E+05& 1.03268E+05& 0.000&0.000\\
                       6.21900 & 0.65507    &1.28368  &  46.2444& 45.3545 & 1.32146E+02& 1.34739E+02& 0.172&0.176\\
                       6.31900 & 0.15270    &1.18405  &  33.8740& 33.2233 & 7.73910E+02& 6.90755E+06& 0.029&0.030\\
                       6.33400 & 0.00002    &1.16928  &  32.2838& 31.6656 & 6.77527E+06& 6.90755E+06& 0.000&0.000\\
                       6.42900 & 0.01233    &1.07429  &  23.3854& 22.9594 & 1.38889E+04& 1.41466E+04& 0.002&0.002\\
                       6.51300 & 0.01217    &0.98992  &  17.1872& 16.8828 & 1.91431E+04& 1.94883E+04& 0.001&0.001\\
                       6.51600 & 0.00006    &0.98675  &  16.9821& 16.6819 & 3.95940E+06& 4.03065E+06& 0.000&0.000\\
                        \hline
\end{tabular}
\end{table}

\clearpage
\begin{table}
    \scriptsize
    \begin{tabular}{|c|c|c|c|c|c|c|c|c|c|}
        \hline$^{100}$Sr\\
        \hline E$_{x}$(MeV) & NME & $Q_{\beta^-}$ (MeV) & $F^{(GM)}_{\beta^{-}}$\cite{Gov71} & $F^{(TW)}_{\beta^{-}}$ & PHL$^{(GM)}$\cite{Gov71} & PHL$^{(TW)}$ &  I$^{(GM)}_{(\beta^-)}$ \cite{Gov71} & I$^{(TW)}_{(\beta^-)}$\\
        \hline         6.57300 & 0.00003    &0.93005  &  13.6192& 13.3789 & 9.25172E+06& 9.41784E+06& 0.000&0.000\\
                       6.58400 & 0.00331    &0.91930  &  13.0440& 12.8128 & 9.25986E+04& 9.42691E+04& 0.000&0.000\\
                       6.70900 & 0.00618    &0.79391  &  7.61027& 7.47817 & 8.50784E+04& 8.65813E+04& 0.000&0.000\\
                       6.72200 & 0.11568    &0.78109  &  7.17285& 7.04917 & 4.82458E+03& 4.90923E+03& 0.005&0.005\\
                       6.77000 & 0.01393    &0.73306  &  5.70097& 5.60592 & 5.04157E+04& 5.12705E+04& 0.000&0.000\\
                       6.89500 & 0.03451    &0.60845  &  2.93440& 2.88738 & 3.95262E+04& 4.01699E+04& 0.001&0.001\\
                       7.08500 & 0.00065    &0.41782  &  0.80160& 0.78904 & 7.62566E+06& 7.74711E+06& 0.000&0.000\\
                       7.10800 & 0.28661    &0.39513  &  0.66388& 0.65345 & 2.10383E+04& 2.13742E+04& 0.000&0.001\\
                       7.25000 & 0.05443    &0.25270  &  0.15179& 0.14949 & 4.84551E+05& 4.92007E+05& 0.000&0.000\\
                       7.30300 & 0.00180    &0.19970  &  7.12037E-02& 7.01109E-02 & 3.12973E+07& 3.17851E+07& 0.000&0.000\\
                       7.34900 & 0.20453    &0.15433  &  3.14666E-02& 3.09831E-03 & 6.21998E+05& 6.31705E+05& 0.000&0.000\\
                       7.43700 & 0.00053    &6.55474E-02  & 2.22245E-03&2.19039E-03  & 3.40802E+09& 3.45789E+09& 0.000&0.000\\
                       7.50100 & 0.00000    &2.44828E-03  &  1.11208E-07&1.39749E-07  & 1.78361E+18& 1.41935E+18& 0.000&0.000\\
                 \hline$^{152}$Nd\\ \hline
                             0.00700&     0.00023&  1.10500& 64.4753& 63.0324& 2.69504E+05& 2.75673E+05& 3.206&3.201\\
                             0.04300&     0.00018&  1.06233& 57.1811& 55.9198& 3.98034E+05& 4.07012E+05& 2.170&2.168\\
                             0.04800&     0.00034&  1.05678& 56.0922& 54.8558& 2.12257E+05& 2.17040E+05& 4.070&4.066\\
                             0.08500&     0.00021&  1.01958& 49.1965& 48.1182& 3.85716E+05& 3.94359E+05& 2.240&2.238\\
                             0.16100&     0.00075&  0.94396& 37.1767& 36.3803& 1.42898E+05& 1.46027E+05& 6.046&6.043\\
                             0.16100&     0.00142&  0.94376& 37.1486& 36.3528& 7.57971E+04& 7.74564E+04& 11.39&11.39\\
                             0.17200&     0.00001&  0.93269& 35.5966& 34.8310& 9.15658E+06& 9.35783E+06& 0.094&0.094\\
                             0.17500&     0.00053&  0.93033& 35.2726& 34.5135& 2.13186E+05& 2.17876E+05& 4.052&4.050\\
                             0.19900&     0.00037&  0.90628& 32.0940& 31.3982& 3.39245E+05& 3.46763E+05& 2.547&2.545\\
                             0.22900&     0.00042&  0.87600& 28.4051& 27.7849& 3.34891E+05& 3.42366E+05& 2.580&2.578\\
                             0.27400&     0.00278&  0.83105& 23.5294& 23.0191& 6.12332E+04& 6.25907E+04& 14.10&14.09\\
                             0.32400&     0.00006&  0.78073& 18.8524& 18.4532& 3.53330E+06& 3.60975E+06& 0.245&0.244\\
                             0.32500&     0.00058&  0.77950& 18.7474& 18.3506& 3.70394E+05& 3.78405E+05& 2.332&2.332\\
                             0.33100&     0.00025&  0.77444& 18.3211& 17.9336& 8.65513E+05& 8.84212E+05& 0.998&0.998\\
                             0.33300&     0.00003&  0.77208& 18.1243& 17.7411& 7.84943E+06& 8.01899E+06& 0.110&0.110\\
                             0.33400&     0.00036&  0.77115& 18.0476& 17.6660& 6.17153E+05& 6.30484E+05& 1.400&1.400\\
                             0.33500&     0.00000&  0.77037& 17.9829& 17.6027& 8.69619E+08& 8.88405E+08& 0.001&0.001\\
                             0.34200&     0.00008&  0.76308& 17.3901& 17.0228& 2.78870E+06& 2.84887E+06& 0.310&0.310\\
                             0.35000&     0.00121&  0.75540& 16.7806& 16.4280& 1.97513E+05& 2.01752E+05& 4.374&4.374\\
                             0.36800&     0.00002&  0.73716& 15.3978& 15.0799& 1.05188E+07& 1.07405E+07& 0.082&0.082\\
                             0.37400&     0.00000&  0.73059& 14.9209& 14.6152& 1.47159E+08& 1.50237E+08& 0.006&0.006\\
                             0.39700&     0.00000&  0.70821& 13.3799& 13.1094& 2.66595E+08& 2.72097E+08& 0.003&0.003\\
                             0.40200&     0.00000&  0.70320& 13.0521& 12.7882& 3.55337E+09& 3.62670E+09& 0.000&0.000\\
                             0.45200&     0.00003&  0.65259& 10.0639& 9.85865& 1.33142E+07& 1.35914E+07& 0.065&0.065\\
                             0.47600&     0.00005&  0.62948& 8.88352& 8.70489& 9.68863E+06& 9.88744E+06& 0.089&0.089\\
                             0.48800&     0.00349&  0.61732& 8.30575& 8.13928& 1.38175E+05& 1.41001E+05& 6.252&6.259\\
                             0.51100&     0.00796&  0.59432& 7.29006& 7.14343& 6.90199E+04& 7.04367E+04& 12.51&12.529\\
                             0.54600&     0.00061&  0.55945& 5.92805& 5.80563& 1.10052E+06& 1.12373E+06& 0.785&0.785\\
                             0.55500&     0.00137&  0.54988& 5.58944& 5.47456& 5.23679E+05& 5.34669E+05& 1.650&1.651\\
                             0.60100&     0.00004&  0.50426& 4.16753& 4.08228& 2.47689E+07& 2.52862E+07& 0.035&0.035\\
                             0.60200&     0.00299&  0.50322& 4.13846& 4.05387& 3.23348E+05& 3.30095E+05& 2.672&2.673\\
                             0.64600&     0.00150&  0.45851& 3.02789& 2.97763& 8.84107E+05& 9.02058E+05& 0.977&0.978\\
                             0.65000&     0.00629&  0.45524& 2.95622& 2.89755& 2.15336E+05& 2.19696E+05& 4.012&4.017\\
                             0.69600&     0.00589&  0.40853& 2.06216& 2.02139& 3.29366E+05& 3.36009E+05& 2.623&2.626\\
                             0.72000&     0.00739&  0.38524& 1.69907& 1.66403& 3.18926E+05& 3.25589E+05& 2.709&2.710\\
                             0.74100&     0.00000&  0.36417& 1.41219& 1.38322& 1.21197E+09& 1.23735E+09& 0.001&0.001\\
                             0.74700&     0.00029&  0.35817& 1.33734& 1.31027& 1.02657E+07& 1.04794E+07& 0.084&0.084\\
                             0.76400&     0.00047&  0.34126& 1.14121& 1.11921& 7.42169E+06& 7.57038E+06& 0.116&0.117\\
                             0.78500&     0.00071&  0.31972& 0.92353& 0.90484& 6.12158E+06& 6.24805E+06& 0.141&0.141\\
                             0.80600&     0.00082&  0.29870& 0.74084& 0.72553& 6.59923E+06& 6.73842E+06& 0.131&0.131\\
                             0.81200&     0.00282&  0.29257& 0.69283& 0.67836& 2.04836E+06& 2.09207E+06& 0.422&0.422\\
                             0.82100&     0.00469&  0.28442& 0.63250& 0.61922& 1.34848E+06& 1.37739E+06& 0.641&0.641\\
                             0.85100&     0.01661&  0.25449& 0.44274& 0.43351& 5.44359E+05& 5.55942E+05& 1.587&1.587\\
                             0.91700&     0.00059&  0.18846& 0.17078& 0.16689& 3.99157E+07& 4.08462E+07& 0.022&0.022\\
                             0.92400&     0.00261&  0.18069& 0.14965& 0.14621& 1.02513E+07& 1.04925E+07& 0.084&0.084\\
                             0.96800&     0.00007&  0.13735& 6.3563E-02& 6.1968E-02& 9.47034E+08& 9.71424E+08 & 0.001&0.001\\
                             0.96800&     0.00093&  0.13673& 6.2679E-02& 6.1101E-02& 6.89166E+07& 7.06975E+07 & 0.013&0.012\\
                             1.00300&     0.00039&  1.0210E-01& 2.5429E-02& 2.4708E-02& 4.00310E+08& 4.11999E+08 & 0.002&0.002\\
                            \hline
\end{tabular}
\end{table}

\clearpage \centering

\end{document}